\documentclass[aps,prl,twocolumn,superscriptaddress,amsmath]{revtex4-2}
\usepackage{graphicx}
\usepackage{hyperref}
\usepackage{mathrsfs}
\usepackage{bm}
\usepackage{color}
\usepackage{siunitx}
\usepackage[capitalize]{cleveref}
\usepackage[T1]{fontenc}
\hypersetup{hypertex=true,
	colorlinks=true,
	anchorcolor=blue,
	linkcolor=blue,
    citecolor=blue,
	urlcolor=blue}
\begin{document}

\title{Spin-textured orbitals in altermagnetic artificial atoms}

\author{Yue Mao}
\affiliation{International Center for Quantum Materials, School of Physics, Peking University, Beijing 100871, China}
\affiliation{Hefei National Laboratory, Hefei 230088, China}
\author{Yu-Chen Zhuang}
\affiliation{International Center for Quantum Materials, School of Physics, Peking University, Beijing 100871, China}
\affiliation{Hefei National Laboratory, Hefei 230088, China}
\author{Cheng-Ming Miao}
\affiliation{International Center for Quantum Materials, School of Physics, Peking University, Beijing 100871, China}
\affiliation{Hefei National Laboratory, Hefei 230088, China}
\author{Yu-Fei Sun}
\affiliation{International Center for Quantum Materials, School of Physics, Peking University, Beijing 100871, China}
\affiliation{Hefei National Laboratory, Hefei 230088, China}
\author{Qing-Feng Sun}
\email[]{sunqf@pku.edu.cn}
\affiliation{International Center for Quantum Materials, School of Physics, Peking University, Beijing 100871, China}
\affiliation{Hefei National Laboratory, Hefei 230088, China}

\date{\today}

\begin{abstract}
Artificial atoms provide a versatile platform for engineering atomic-like orbitals, yet spin generally remains a passive degree of freedom in their orbital structure.
Here, we introduce the concept of altermagnetic artificial atoms formed by confining electrons with momentum-dependent spin splitting.
We show that altermagnetism reconstructs conventional confined orbitals into spin-textured orbitals, with spatially distinct distributions of opposite spin components.
The resulting confined spectrum retains a twofold degeneracy protected by the combined $C_{4z}\mathcal{T}$ symmetry.
These spin textures persist in higher-energy states, where additional radial structures combine with the characteristic angular spin pattern.
Furthermore, strain resolves the degenerate orbital pairs into spin-polarized states, and continuously tunes their energy splitting.
Our results establish altermagnetic artificial atoms as a route to engineering spin-dependent orbital structures in quantum-confined systems.
\end{abstract}

\maketitle

\textit{Introduction}---Quantum confinement converts a continuous electronic spectrum into discrete levels, giving rise to artificial atoms with atomic-like orbitals \cite{crommie_confinement_1993,ashoori_electrons_1996,kouwenhoven_fewelectron_2001,reimann_electronic_2002,buluta_natural_2011}.
Beyond reproducing natural atoms, artificial atoms provide a versatile platform for engineering quantum states through artificial design.
Their energy levels and wavefunctions can be controlled through the size and shape of the confinement, while their comparatively large spatial extent allows direct real-space imaging of the confined orbitals \cite{crommie_confinement_1993,lee_imaging_2016,stilp_very_2021,ge_direct_2024,mao_orbital_2025}.
Moreover, coupling between artificial atoms has enabled artificial molecular states \cite{schedelbeck_coupled_1997,oosterkamp_microwave_1998,bayer_coupling_2001,fu_relativistic_2020,dou_highyield_2023,zheng_molecular_2023,ge_giant_2023,zhou_relativistic_2024}, whereas the confinement of Dirac electrons in graphene has produced relativistic artificial orbitals \cite{silvestrov_quantum_2007,matulis_quasibound_2008,wang_observing_2013,zhao_creating_2015,mao_realization_2016,ghahari_berry_2017,zheng_coexistence_2022,ge_direct_2024,zhuang_atomic_2025}.
However, conventional confinement acts solely on the charge degree of freedom and does not couple to spin.
Consequently, spin serves as an internal label of the confined states rather than determining their orbital wavefunctions.

Altermagnetism has recently emerged as a distinct form of magnetic order.
In altermagnets, the interplay of compensated magnetic order and crystal symmetry produces momentum-dependent spin splitting in the electronic bands despite vanishing net magnetization \cite{smejkal_emerging_2022,smejkal_chiral_2023,bai_altermagnetism_2024,liu_twisted_2024,kaushal_altermagnetism_2025,jungwirth_symmetry_2026,li_marginal_2026,zhu_design_2025,jin_interactiondriven_2026}.
This unconventional magnetism has stimulated extensive studies of electronic structures
\cite{fedchenko_observation_2024,krempasky_altermagnetic_2024,osumi_observation_2024,reimers_direct_2024,wei_gapless_2024,jiang_metallic_2025,zhang_crystalsymmetrypaired_2025,zhu_altermagnetic_2026} and transport phenomena \cite{gonzalez-hernandez_efficient_2021,ouassou_dc_2023,banerjee_altermagnetic_2024,lu_josephson_2024,cheng_fieldfree_2024,zhou_crystal_2024,chakraborty_perfect_2025,nagae_spinpolarized_2025,li_spinpolarized_2026,monkman_persistent_2026,yi_spin_2026}.
It has also opened routes toward topological states \cite{ghorashi_altermagnetic_2024,ezawa_detecting_2024,li_altermagnetisminduced_2025,qu_altermagnetic_2025,sun_altermagnetisminduced_2025,wan_helical_2025,fukaya_crossed_2026} and diverse control schemes and device functionalities \cite{smejkal_giant_2022,amin_nanoscale_2024,sun_tunneling_2025,zhang_theory_2025,zhou_manipulation_2025,ding_ferroelastically_2025,duan_antiferroelectric_2025,gu_ferroelectric_2025,liu_altermagnetic_2026}.
A parallel question then arises for quantum-confined systems: can altermagnetism extend the design principles of artificial atoms by encoding spin into confined orbital degrees of freedom?

\begin{figure}[]
	\includegraphics[width=\columnwidth]{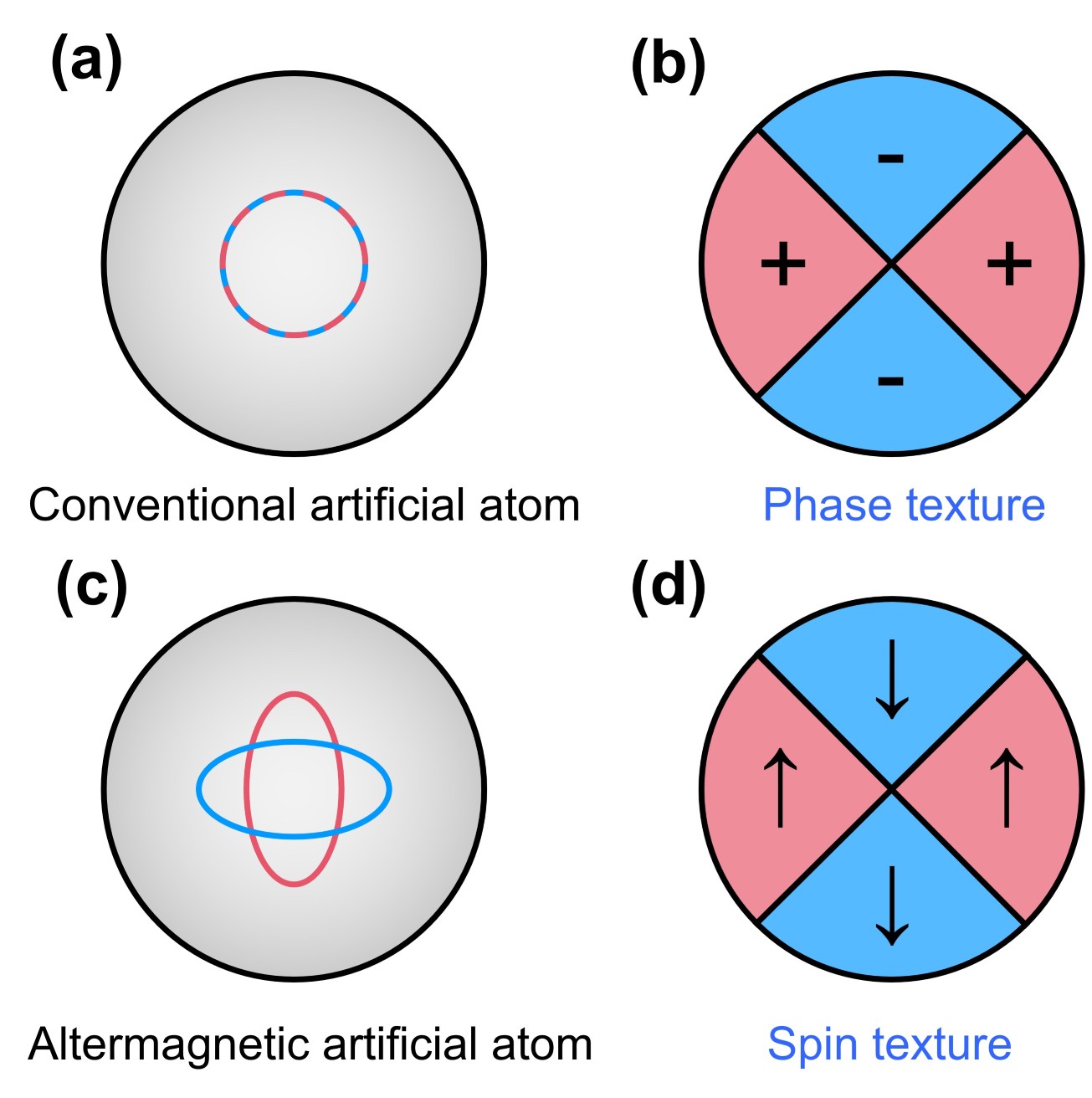}
	\centering
	\caption{Concept of altermagnetic artificial atoms.
    (a) A conventional artificial atom formed by confining a spin-degenerate system.
    (b) Phase texture of a conventional $d$-orbital, represented by alternating wavefunction signs in a real orbital representation.
    (c) An altermagnetic artificial atom formed by confining a system with momentum-dependent spin splitting.
    The inner constant-energy contours are spin degenerate in (a) and spin split in (c).
    (d) Spin texture of a $d$-wave altermagnetic artificial orbital pair, formed by spatially alternating opposite-spin components.}
	\label{FIG1}
\end{figure}

In this Letter, we introduce the concept of altermagnetic artificial atoms by incorporating altermagnetic electronic structures into quantum-confined systems, and demonstrate the emergence of spin-textured orbitals.
We show that the spin-dependent anisotropy of altermagnetism transforms conventional spin-degenerate orbitals into quantum states with intertwined spin and orbital degrees of freedom.
These orbitals exhibit spatially alternating spin textures that are related by the symmetries of the altermagnetism.
We further identify higher-energy counterparts with additional radial structure and related spin textures.
Finally, we show that strain lifts the degeneracy between the paired spin-textured orbitals and continuously tunes their energy splitting.

\textit{Altermagnetic artificial atoms}---In conventional artificial atoms [Fig. \ref{FIG1}(a)], quantum confinement discretizes the continuous electronic spectrum and gives rise to a series of atomic-like orbitals.
For rotationally symmetric confinement, these orbitals carry well-defined angular momenta, encoded in the phase accumulated by their wavefunctions upon encircling the confinement center.
In a real orbital representation, this phase structure appears as alternating wavefunction signs around the confinement center, forming characteristic orbital phase textures [e.g., the $d$-orbital phase texture shown in Fig. \ref{FIG1}(b)].
Spin, however, remains decoupled from orbital motion, leaving the confined orbitals spin degenerate.

Altermagnetic artificial atoms offer a distinct route to orbital design [Fig. \ref{FIG1}(c)].
Owing to momentum-dependent spin splitting, electrons with opposite spins favor different directions of motion in the parent altermagnetic system.
Under quantum confinement, this spin-dependent motion becomes encoded in the spatial orbital structure, so that spin no longer remains a passive label.
The resulting spin-textured artificial orbitals exhibit spatially alternating spin-up and spin-down contributions [Fig. \ref{FIG1}(d)], analogous to the alternating wavefunction signs underlying the phase textures of conventional orbitals [Fig. \ref{FIG1}(b)].

To establish the above picture, we describe altermagnetic artificial atoms using a representative low-energy Hamiltonian
\begin{equation}
	H=t_0(\hat{k}_x^2+\hat{k}_y^2)+t_J(\hat{k}_x^2-\hat{k}_y^2)\sigma_z+V(r).
\label{Eq1}
\end{equation}
Here, $\hat{k}_x=-i\partial_x, \hat{k}_y=-i\partial_y$ denote the wavevector operators, $t_0=1$ sets the kinetic-energy scale, $t_J$ is the altermagnetic strength responsible for momentum-dependent spin splitting \cite{smejkal_giant_2022,li_marginal_2026,lu_josephson_2024}, $V(r)$ is an isotropic hard-wall confining potential, and $\sigma$ is the Pauli matrix in spin space.
The conventional artificial atom is recovered for $t_J=0$, whereas $t_J\neq0$ describes its altermagnetic counterpart.
Owing to the rotational symmetry of the kinetic and confinement terms, we choose the principal axes of the altermagnetic order as the $x$ and $y$ axes without loss of generality.
We discretize the Hamiltonian on a square lattice and numerically calculate the confined energy levels and wavefunctions, with details provided in the End Matter.

\begin{figure*}[ht]
	\includegraphics[width=1.8\columnwidth]{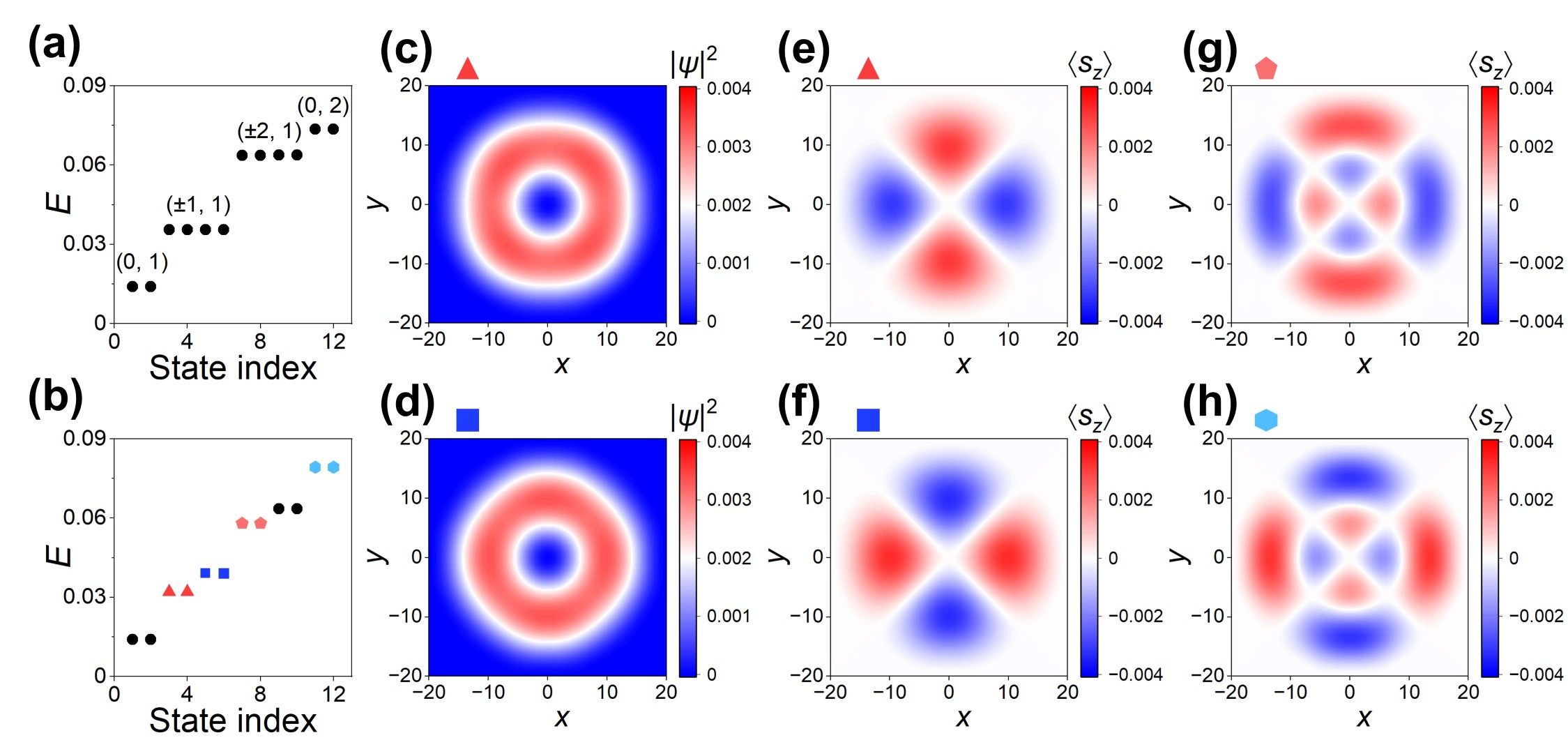}
	\centering
	\caption{Formation of spin-textured orbitals in altermagnetic artificial atoms.
    (a) Energy spectrum of a conventional artificial atom with $t_J=0$, where states are labeled by angular-momentum and principal quantum numbers $(m,n)$.
    (b) Energy spectrum of the altermagnetic artificial atom with $t_J=0.2$. Colored symbols mark representative degenerate pairs.
    (c),(d) Probability densities of the two degenerate pairs split from the $(\pm1,1)$ states, summed over each pair.
    (e),(f) Corresponding local spin densities.
	(g),(h) Local spin densities of confined states marked in (b) with additional radial structure.}
	\label{FIG2}
\end{figure*}

\textit{Formation and symmetry of spin-textured orbitals}---For conventional artificial atoms with $t_J=0$, we calculate the low-energy confined-state spectrum, as shown in Fig. \ref{FIG2}(a).
To understand the structure of this spectrum, we consider the analytical solution for a hard-wall confinement.
An orbital characterized by the angular momentum quantum number $m$ and principal quantum number $n$ has the wavefunction $\psi_{m,n} \propto J_{|m|}\left(x_{|m|,n}r/R\right)e^{im\phi}$, with the corresponding energy $E_{m,n}=t_0x_{|m|,n}^2/R^2$ \cite{crommie_confinement_1993}.
Here, $J_{|m|}$ is the $|m|$-order Bessel function of the first kind, $x_{|m|,n}$ denotes its $n$-th zero, $r$ is the distance from the confinement center, $\phi$ is the polar angle, and $R$ is the confinement radius.
The ordering of the confined-state energies is therefore determined by the Bessel-function zeros, giving the first few states as $(m,n)=(0,1), (\pm1,1), (\pm2,1)$, and $(0,2)$, labelled in Fig. \ref{FIG2}(a).
Owing to spin degeneracy and the degeneracy between positive and negative angular momenta, these states have degeneracies of $2$, $4$, $4$, and $2$, respectively.
The numerical spectrum closely reproduces this analytical level structure.

For altermagnetic artificial atoms, the altermagnetic term reconstructs the original orbital states and splits the fourfold degeneracy into two doubly degenerate levels, as shown in Fig. \ref{FIG2}(b).
A representative example is the splitting of the original $(\pm1,1)$ states, which correspond to spin-degenerate $p$ orbitals in conventional artificial atoms.
The probability densities of the two resulting orbital pairs are shown in Figs. \ref{FIG2}(c), \ref{FIG2}(d), together with their corresponding local spin densities in Figs. \ref{FIG2}(e), \ref{FIG2}(f).
Although the probability densities retain the ring-like profile of the original $p$ orbitals, described by $J^2_1(x_{1,1}r/R)$, their local spin densities develop a spatially alternating pattern. For the lower-energy pair, the local spin density is negative along the $x$ axis and positive along the $y$ axis, corresponding to enhanced spin-down and spin-up contributions, respectively. The higher-energy pair exhibits the opposite arrangement, with the signs of the local spin density interchanged between the two axes. The two orbital pairs therefore carry opposite spin textures.

Altermagnetism reconstructs the conventional $p$ orbitals through the angular structure of the $(k_x^2-k_y^2)\sigma_z$ term.
Since $k_x^2-k_y^2$ carries angular-momentum components with $\Delta m=\pm2$, it directly couples the originally degenerate $m=+1$ and $m=-1$ states.
The resulting eigenstates are therefore reorganized into the $p_x$ and $p_y$ orbitals, with $p_x\propto\psi_{1,1}+\psi_{-1,1}$ and $p_y\propto(\psi_{1,1}-\psi_{-1,1})$, spatially oriented along the $x$ and $y$ axes, respectively.
For each spin sector, this coupling lifts the degeneracy between the $p_x$ and $p_y$ orbitals.
Because the altermagnetic term has opposite signs for spin-up and spin-down electrons, the $p_x$-$p_y$ level ordering is reversed between the two spin sectors.
Consequently, the remaining degenerate pairs are $(p_y,\uparrow)$ with $(p_x,\downarrow)$ and $(p_x,\uparrow)$ with $(p_y,\downarrow)$, consistent with the spin textures shown in Figs. \ref{FIG2}(e), \ref{FIG2}(f).

The twofold degeneracy of the altermagnetic artificial-atom spectrum in Fig. \ref{FIG2}(b) is protected by the combined symmetry $C_{4z}\mathcal{T}$, which satisfies $(C_{4z}\mathcal{T})H(C_{4z}\mathcal{T})^\dagger=H$.
Here, $C_{4z}$ represents a rotation by $\pi/2$ around the $z$ axis, and $\mathcal{T}$ denotes the time-reversal operator.
Since the Hamiltonian commutes with $\sigma_z$, the confined states can be chosen as eigenstates of $\sigma_z$.
The operation $C_{4z}\mathcal{T}$ reverses the spin and maps each state onto a distinct state at the same energy.
Each confined state therefore has an opposite-spin degenerate partner, accounting for the twofold degeneracy throughout the spectrum in Fig. \ref{FIG2}(b).
For the $p$ orbitals discussed above, $C_{4z}\mathcal{T}$ transforms $(p_x,\sigma)$ into $(p_y,-\sigma)$, thereby enforcing $E_{p_x,\sigma}=E_{p_y,-\sigma}$.

The spin textures of the confined orbitals exhibit a striking analogy to the spatial phase structures of $d$ orbitals in both artificial and natural atoms.
For conventional $d$ orbitals, the angular dependence associated with $|m|=2$ gives rise, in a real orbital representation, to wavefunction signs that are identical along opposite directions but reversed along orthogonal directions [Fig. \ref{FIG1}(b)].
The altermagnetic artificial orbitals exhibit an analogous $d$-wave-like structure in their local spin densities: the spin polarization has the same sign along opposite directions but reverses between perpendicular directions [see Figs. \ref{FIG1}(d), \ref{FIG2}(e), \ref{FIG2}(f)].
This correspondence establishes a spin-space counterpart of the characteristic phase textures of conventional orbitals.

\begin{figure*}[]
	\includegraphics[width=1.8\columnwidth]{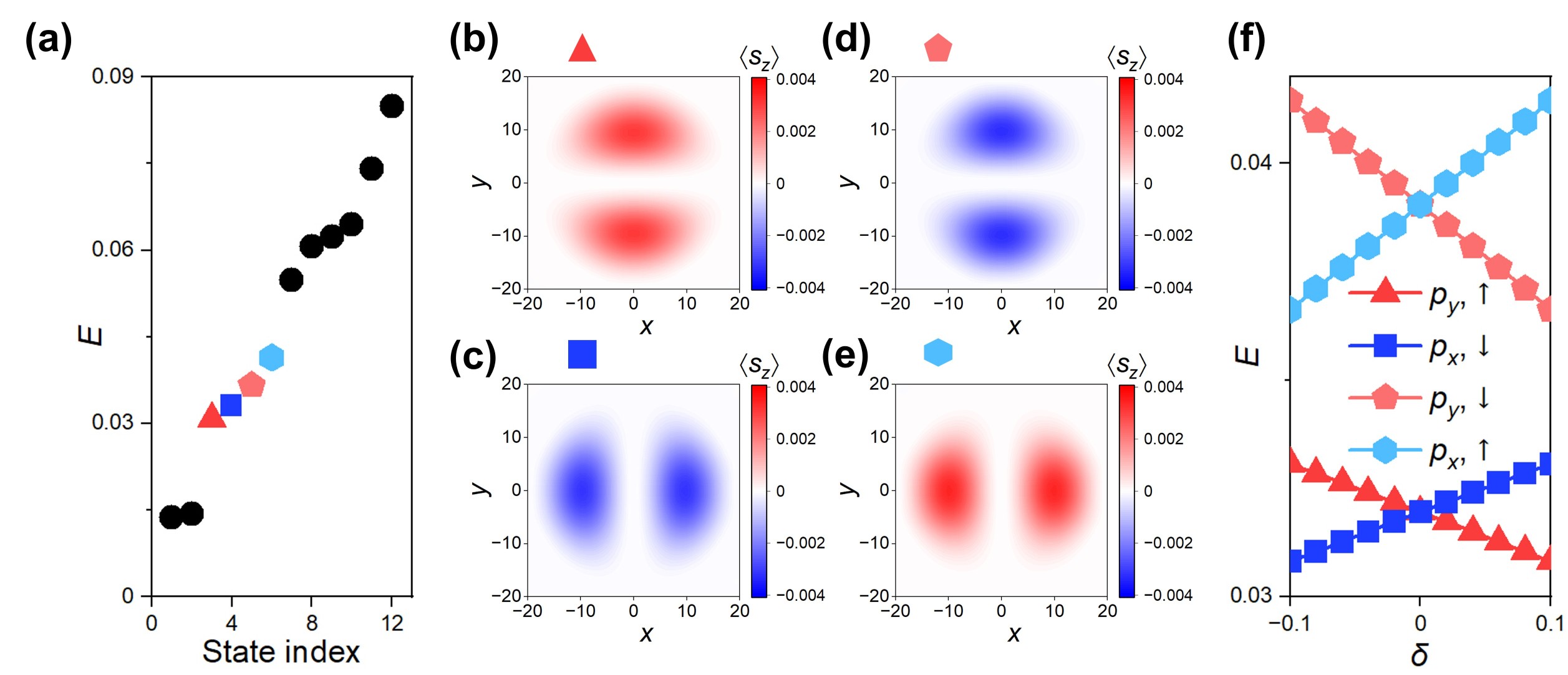}
	\centering
	\caption{Strain control of orbital degeneracy in altermagnetic artificial atoms.
    (a) Energy spectrum of the altermagnetic artificial atom at $\delta=0.1$, with colored symbols indicating representative split states.
    (b)-(e) Local spin densities of the states marked by colored symbols in (a).
        (f) Strain dependence of the energies of the four $p$ orbital states.}
	\label{FIG3}
\end{figure*}

The spin textures of altermagnetic confined orbitals are not limited to the simple states discussed above, but also emerge in higher-energy states with additional radial structure.
Figures \ref{FIG2}(g), \ref{FIG2}(h) present representative examples from the higher-energy levels marked in Fig. \ref{FIG2}(b).
Similar to the lower-energy states, two pairs of orbitals with opposite spin textures are formed.
In addition to the alternating local spin density along the polar direction, these higher-energy orbitals exhibit an additional sign reversal along the radial direction, with opposite signs in the inner and outer regions.
This radial spin reversal resembles the behavior of a confined $d$ orbital with an internal radial node.
For such a conventional orbital, the wavefunction changes sign across the internal radial node.
Here in the altermagnetic confined orbitals, this additional radial structure originates from the orbital reconstruction between the $(\pm2,1)$ and $(0,2)$ states.
Since the altermagnetic term $k_x^2-k_y^2$ carries angular-momentum components with $\Delta m=\pm2$, it couples the $(0,2)$ state to the $(\pm2,1)$ states.
Their close energies further enhance this coupling [see Fig. \ref{FIG2}(a)].
The $(0,2)$ state has one internal radial node, whereas the $(\pm2,1)$ states have none; the latter instead carry the $|m|=2$ angular structure.
Their reorganization thus combines the radial nodal structure with the $d$-wave-like angular character, underlying the alternating local spin density along both the radial and polar directions [Figs. \ref{FIG2}(g), \ref{FIG2}(h)].

\textit{Strain control of orbital degeneracy}---The symmetry-protected orbital degeneracy provides a natural handle for external control.
Strain can break the rotational symmetry of altermagnets \cite{belashchenko_giant_2025,huang_spin_2025,cao_symmetry_2025}, providing a continuously tunable route to control the orbital degeneracy in altermagnetic artificial atoms.
A uniaxial strain along the $x$ direction modifies the electronic motion through changes in the interatomic distances and orbital overlaps, producing anisotropic responses along the $x$ and $y$ directions.
We phenomenologically parameterize this anisotropy by $k_x^2\rightarrow(1+\delta)k_x^2$ and $k_y^2\rightarrow(1-\delta)k_y^2$, where $\delta$ characterizes the strain-induced anisotropy.
The symmetric form is adopted as a convenient parametrization of the anisotropic response, with the overall change in the energy scale absorbed into the model parameters.

We next calculate the confined-state spectrum of altermagnetic artificial atoms under strain.
As shown in Fig. \ref{FIG3}(a), strain lifts the twofold degeneracy, splitting the two degenerate $p$-orbital pairs into four nondegenerate states.
The strain-induced splitting separates the originally degenerate states into distinct spin-polarized orbitals.
Consequently, the spatially alternating local spin density at a given energy is resolved into separate $p_x$ and $p_y$ orbitals with opposite spin polarizations, as shown in Figs. \ref{FIG3}(b)-\ref{FIG3}(e).
The strain-induced enhancement of the kinetic energy along the $x$ direction dominantly shifts each $p_x$ state upward relative to its originally degenerate $p_y$ partner, yielding the energy ordering $(p_y,\uparrow)$, $(p_x,\downarrow)$, $(p_y,\downarrow)$, and $(p_x,\uparrow)$ from low to high.

We further investigate the evolution of the confined states with strain. Figure \ref{FIG3}(f) shows the strain dependence of their energies. As the strain parameter $\delta$ increases, the energies of the two $p_x$ orbitals increase, whereas those of the two $p_y$ orbitals decrease. Although the strained Hamiltonian $H(\delta)$ no longer preserves the $C_{4z}\mathcal{T}$ symmetry for $\delta\neq0$, the Hamiltonians at opposite values of $\delta$ are related by this operation:
\begin{equation}
    (C_{4z}\mathcal{T})H(\delta)(C_{4z}\mathcal{T})^\dagger=H(-\delta).
\end{equation}
Accordingly, $C_{4z}\mathcal{T}$ maps a $(p_x,\sigma)$ state at $\delta$ onto a $(p_y,-\sigma)$ state at $-\delta$, leading to
\begin{equation}
    E_{p_x,\sigma}(\delta)=E_{p_y,-\sigma}(-\delta).
\end{equation}
This relation accounts for the symmetric evolution with respect to $\delta=0$ in Fig. \ref{FIG3}(f).

\begin{figure}[t]
	\includegraphics[width=1.0\columnwidth]{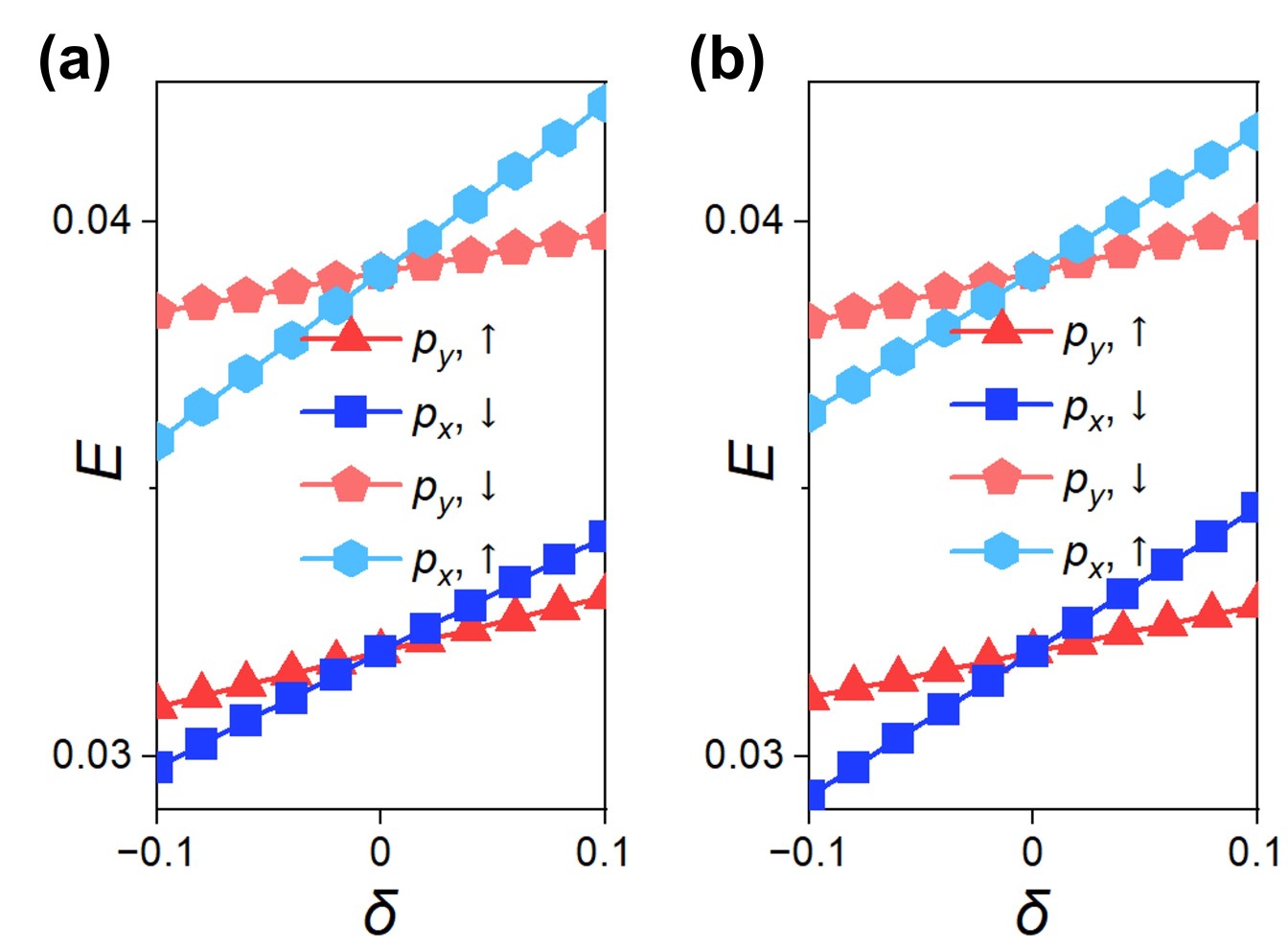}
	\centering
	\caption{Strain dependence of the energies of the four $p$ states.
    (a) Strain is applied only along the $x$ direction, to both the kinetic and altermagnetic terms.
    (b) Strain is applied only along the $x$ direction to the kinetic term, while the altermagnetic term remains unchanged.}
	\label{FIG4}
\end{figure}

The orbital splitting is robust against the specific implementation of strain.
In the above analysis, we adopt the symmetric modification $k_x^2\rightarrow(1+\delta)k_x^2$ and $k_y^2\rightarrow(1-\delta)k_y^2$, with opposite modulations along the two directions.
The orbital splitting is also obtained when only the $x$ direction is modified, $k_x^2\rightarrow(1+\delta)k_x^2$, as shown in Fig. \ref{FIG4}(a) in the End Matter.
We further consider the strain response of the altermagnetic term. Given the diverse microscopic origins of altermagnetism \cite{smejkal_emerging_2022,jungwirth_symmetry_2026}, the altermagnetic term can exhibit different responses to strain in different systems.
The treatment considered above, in which strain modifies both the kinetic and altermagnetic terms, represents one limiting case corresponding to a strain-sensitive altermagnetic contribution.
We further consider the opposite limit, corresponding to a strain-insensitive altermagnetic contribution, in which the strain-induced modification acts only on the kinetic term.
We find that a similar orbital splitting is obtained in this case, as shown in Fig. \ref{FIG4}(b) in the End Matter.
The persistence of the orbital splitting under these different strain implementations reflects the same underlying principle: each of them breaks the $C_{4z}\mathcal{T}$ symmetry that protects the twofold degeneracy, thereby allowing the originally degenerate confined states to split.

\textit{Conclusions and outlook}---We introduce the concept of altermagnetic artificial atoms, in which quantum confinement converts the spin-dependent electronic motion of altermagnets into confined orbital states with characteristic spin textures.
Unlike conventional artificial atoms, where spin remains a passive label of the orbital states, altermagnetic confinement encodes spin into their spatial orbital structure.
We show that the confined spectrum retains a twofold degeneracy protected by the combined $C_{4z}\mathcal{T}$ symmetry, with the paired orbitals exhibiting spatially alternating local spin densities.
Similar spin textures also emerge in higher-energy states with additional radial structure.
Furthermore, we find that strain breaks the $C_{4z}\mathcal{T}$ symmetry, resolves the degenerate states into spin-polarized orbitals, and continuously tunes their energy splitting.
These results establish altermagnetic artificial atoms as a route to introducing spin-dependent orbital structures into quantum-confined systems, extending the design principles of artificial atoms from spatial orbital control to spin-dependent orbital engineering.

\textit{Acknowledgments}---This work was financially supported by the National Natural Science Foundation of China (Grants No. 12447146, No. 12447147, No. 125B2065, and No. 12374034),
the National Key R and D Program of China (Grant No. 2024YFA1409002),
the Quantum Science and Technology-National Science and Technology Major Project (Grant No. 2021ZD0302403),
the China National Postdoctoral Program for Innovative Talents (Grant No. BX20250182),
the China Postdoctoral Science Foundation (Grants No. 2026M793656, No. 2025T180938 and No. 2024M760070),
and the Postdoctoral Fellowship Program of CPSF (Grant No. GZB20240031).
The computational resources were supported by High-performance Computing Platform of Peking University.

\bibliography{my_AMQD_ref_20260729}

\begin{onecolumngrid}
\subsection{\large End Matter}
\end{onecolumngrid}

\setcounter{equation}{0}
\renewcommand{\theequation}{A\arabic{equation}}

\twocolumngrid

\textit{Computation method}---We start from the continuum Hamiltonian [Eq. (\ref{Eq1}) in the main text]
\begin{equation}
H=t_0(\hat{k}_x^2+\hat{k}_y^2)
+t_J(\hat{k}_x^2-\hat{k}_y^2)\sigma_z+V(r),
\end{equation}
where $t_0$ and $t_J$ characterize the kinetic and altermagnetic terms, respectively. To incorporate different implementations of strain in a unified form, we write the strained Hamiltonian as
\begin{equation}
H(\delta)=t_x\hat{k}_x^2+t_y\hat{k}_y^2
+\left(t_{Jx}\hat{k}_x^2-t_{Jy}\hat{k}_y^2\right)\sigma_z
+V(r).
\end{equation}
For the symmetric strain parametrization used in the main calculations, where both the kinetic and altermagnetic terms are modified along the two directions,
\begin{equation}
\begin{aligned}
t_x&=t_0(1+\delta),\qquad t_y=t_0(1-\delta),\\
t_{Jx}&=t_J(1+\delta),\qquad t_{Jy}=t_J(1-\delta).
\end{aligned}
\end{equation}
For the strain implementation in which only the $x$ direction is modified in both terms,
\begin{equation}
t_x=t_0(1+\delta),\quad
t_y=t_0,\quad
t_{Jx}=t_J(1+\delta),\quad
t_{Jy}=t_J.
\end{equation}
Finally, when the altermagnetic term is taken to be insensitive to strain and the strain-induced modification is applied only to the kinetic term along the $x$ direction, the parameters become
\begin{equation}
t_x=t_0(1+\delta),\quad
t_y=t_0,\quad
t_{Jx}=t_J,\quad
t_{Jy}=t_J.
\end{equation}

We discretize the continuum Hamiltonian on a square lattice using the differential operators.
With the lattice constant set to unity, the resulting lattice Hamiltonian can be written as
\begin{equation}
\begin{aligned}
H_{\rm lat}
=&\sum_{\mathbf n} c_{\mathbf n}^\dagger
\left[h_0+V(\mathbf r_{\mathbf n})\sigma_0\right]c_{\mathbf n}\\
&+\sum_{\mathbf n}
\left(
c_{\mathbf n}^\dagger h_x c_{\mathbf n+\hat{\mathbf x}}
+c_{\mathbf n}^\dagger h_y c_{\mathbf n+\hat{\mathbf y}}
+\mathrm{H.c.}
\right).
\end{aligned}
\end{equation}
Here, $c_{\mathbf n}=(c_{\mathbf n\uparrow},c_{\mathbf n\downarrow})^{T}$,
$\mathbf n$ labels the lattice sites, and $\hat{\mathbf x}$ and
$\hat{\mathbf y}$ denote the unit lattice vectors along the $x$ and $y$ directions, respectively.
The matrices entering the lattice Hamiltonian are
\begin{align}
h_0&=2(t_x+t_y)\sigma_0
+2(t_{Jx}-t_{Jy})\sigma_z,\\
h_x&=-t_x\sigma_0-t_{Jx}\sigma_z,\\
h_y&=-t_y\sigma_0+t_{Jy}\sigma_z.
\end{align}

The circular confinement is implemented by
\begin{equation}
V(r)=
\begin{cases}
0, & r<R,\\
V_0, & r\geq R,
\end{cases}
\end{equation}
with $R=20$ and $V_0=100t_0$. The system is discretized on a $41\times41$ square lattice, and the large value of $V_0$ effectively realizes circular hard-wall confinement. The confined-state energies and wavefunctions are obtained by direct diagonalization of the lattice Hamiltonian.

For an individual confined state $\psi$, the probability density is given by the wavefunction modulus squared,
\begin{equation}
|\psi(\mathbf r)|^2
=
|\psi_{\uparrow}(\mathbf r)|^2
+
|\psi_{\downarrow}(\mathbf r)|^2,
\end{equation}
and the local spin density is
\begin{equation}
\langle s_z(\mathbf r)\rangle
=
\psi^\dagger(\mathbf r)\sigma_z\psi(\mathbf r)
=
|\psi_{\uparrow}(\mathbf r)|^2
-
|\psi_{\downarrow}(\mathbf r)|^2.
\end{equation}
For degenerate states at the same energy, the probability densities and local spin densities are summed over the degenerate states, as indicated by the same colored symbol in the energy spectra.

\end{document}